\documentclass[11pt,a4paper]{article}

\usepackage[a4paper,top=18mm,bottom=17mm,left=24mm,right=24mm,headheight=22pt,headsep=5pt,footskip=14mm]{geometry}
\usepackage[T1]{fontenc}
\usepackage[utf8]{inputenc}
\usepackage{newtxtext,newtxmath}
\usepackage[scaled=0.92]{helvet}
\usepackage{graphicx}
\usepackage{amsmath}
\usepackage{booktabs,tabularx,array,longtable}
\usepackage{enumitem}
\usepackage{xcolor}
\usepackage{caption}
\usepackage{titlesec}
\usepackage{fancyhdr}
\usepackage{microtype}
\usepackage{ragged2e}
\usepackage{float}
\usepackage{url}
\usepackage{cite}
\usepackage{etoolbox}
\usepackage[hidelinks,hyperfootnotes=false]{hyperref}

\definecolor{AGBlue}{HTML}{004CFF}
\definecolor{HeaderGray}{HTML}{606A7A}
\definecolor{RuleGray}{HTML}{D9DEE8}

\hypersetup{
  pdftitle={Improving Requirements Classification with SMOTE-Tomek Preprocessing},
  pdfauthor={Barak Or},
  pdfsubject={Requirements classification under severe class imbalance},
  pdfkeywords={Requirements Engineering, Class Imbalance, SMOTE-Tomek, PROMISE Dataset, Text Classification}
}

\setlist[itemize]{topsep=0.18em,itemsep=0.10em,parsep=0pt,partopsep=0pt}
\setlist[enumerate]{topsep=0.18em,itemsep=0.12em,parsep=0pt,partopsep=0pt}
\setlist[description]{topsep=0.15em,itemsep=0.12em,parsep=0pt,partopsep=0pt}
\AtBeginEnvironment{thebibliography}{\scriptsize\setlength{\itemsep}{0pt}\setlength{\parsep}{0pt}\setlength{\parskip}{0pt}}

\titleformat{\section}
  {\sffamily\bfseries\color{AGBlue}\large}
  {\thesection}{0.6em}{}
\titlespacing*{\section}{0pt}{0.78em}{0.22em}

\titleformat{\subsection}
  {\sffamily\bfseries\normalsize}
  {\thesubsection}{0.6em}{}
\titlespacing*{\subsection}{0pt}{0.62em}{0.14em}

\renewcommand{\headrulewidth}{0.35pt}
\renewcommand{\headrule}{\hbox to\headwidth{\color{RuleGray}\leaders\hrule height \headrulewidth\hfill}}

\newcolumntype{Y}{>{\RaggedRight\arraybackslash}X}
\newcolumntype{L}[1]{>{\RaggedRight\arraybackslash}p{#1}}
\newcommand{\pmv}[2]{#1 $\pm$ #2}
\newcommand{\blfootnote}[1]{%
  \begingroup
  \renewcommand{\thefootnote}{}%
  \footnotetext{#1}%
  \addtocounter{footnote}{-1}%
  \endgroup
}

\newcounter{algorithmbox}
\newenvironment{algorithmbox}[1]{%
  \refstepcounter{algorithmbox}%
  \par\smallskip\noindent
  \begin{minipage}{\textwidth}
  \hrule\vspace{0.24em}
  \textbf{Algorithm \thealgorithmbox}\quad #1
  \vspace{0.18em}\hrule\vspace{0.28em}
}{%
  \vspace{0.24em}\hrule
  \end{minipage}\par\smallskip
}

\begin{document}
\raggedbottom
\thispagestyle{empty}

\noindent\includegraphics[width=6.2cm]{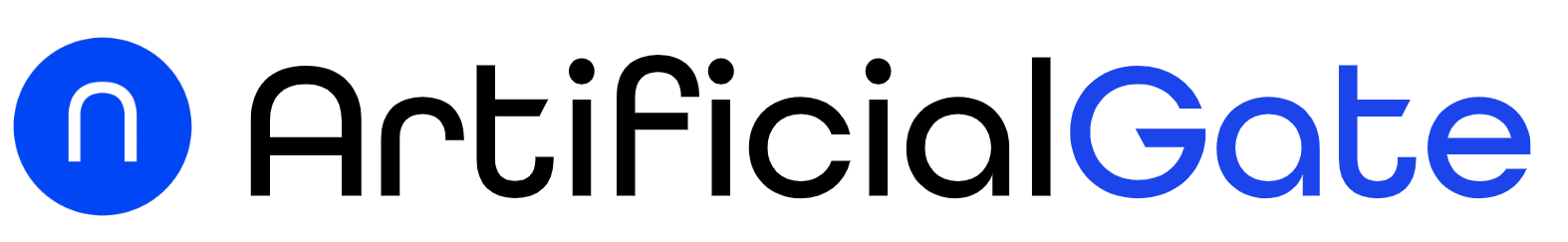}
\vspace{0.22cm}

\noindent{\color{AGBlue}\rule{\textwidth}{1.1pt}}
\vspace{0.34cm}

{\sffamily\bfseries\color{HeaderGray}ArtificialGate Ltd. Research Preprint\par}
\vspace{0.28cm}

{\sffamily\bfseries\fontsize{17}{20}\selectfont\raggedright
Improving Requirements Classification with\\[-0.08em]
SMOTE\mbox{-}Tomek Preprocessing\par}
\vspace{0.38cm}

{\large\bfseries Barak Or\par}
ArtificialGate Ltd., Israel\par
\texttt{barak@artificialgate.ai}\par
\vspace{0.48cm}

{\centering\sffamily\bfseries Abstract\par}
\vspace{0.12em}
{\small
This study emphasizes the domain of requirements engineering by applying the SMOTE-Tomek preprocessing technique, combined with stratified K-fold cross-validation, to address class imbalance in the PROMISE dataset. This dataset comprises 969 categorized requirements, classified into functional and non-functional types. The proposed approach enhances the representation of minority classes while maintaining the integrity of validation folds, leading to a notable improvement in classification accuracy. Logistic Regression achieved \pmv{76.16\%}{2.58\%}, substantially exceeding its baseline of \pmv{59.85\%}{2.52\%}. These results highlight the applicability and efficiency of machine learning models as scalable and interpretable solutions.\par}
\vspace{0.28em}

{\small
\noindent{\sffamily\bfseries\color{AGBlue}Keywords.}\quad
Requirements Engineering; Class Imbalance; SMOTE-Tomek; Imbalanced Datasets; Requirements Classification; Natural Language Processing; Synthetic Oversampling; Text-Based Classification; Supervised Learning\par}

\blfootnote{This research was conducted under ArtificialGate Ltd. The author serves as Academic Director at the Google \& Reichman Tech School, and as an External Lecturer at Technion - Israel Institute of Technology and Reichman University.}

\section{Introduction}
Requirements engineering (RE) is a cornerstone of the software development lifecycle, translating stakeholder needs into actionable system specifications \cite{ref1,ref2,ref3}. Accurate and efficient requirement classification is essential to ensure project clarity, prioritization, and goal alignment. Requirements are typically categorized into functional, non-functional, and subtypes \cite{ref4,ref5,ref6,ref7,ref8}. A survey revealed that over 60\% of failed projects neglected non-functional requirements, underscoring the critical importance of effective classification \cite{ref9}. Despite its significance, requirement classification remains a largely manual process, prone to inconsistencies, inefficiencies, and scalability challenges \cite{ref10}.

Over the years, researchers have explored various approaches to automate requirement classification. Early rule-based systems were among the first attempts, offering interpretable but inflexible solutions that were labor-intensive to maintain \cite{ref11}. The advent of machine learning (ML) brought more adaptive models, demonstrating moderate success, particularly in structured datasets \cite{ref12,ref4,ref7}. However, these methods often face challenges with class imbalance, a prevalent issue in real-world datasets, including the widely used PROMISE dataset for requirements engineering \cite{ref13}. This imbalance skews model predictions toward majority classes, undermining performance on minority classes.

Over the past decade, deep learning (DL) has achieved transformative breakthroughs across multiple domains, including natural language processing (NLP), computer vision, and time series analysis. These advancements are primarily driven by the unprecedented representational power of deep neural networks, which leverage millions, and in some cases billions, of trainable parameters to extract and model intricate, high-dimensional patterns from large-scale datasets \cite{ref14}.

Equally significant are the advances in time series analysis, where DL models have demonstrated remarkable capabilities in tasks such as motion sensing classification and physical quantity estimation \cite{ref15,ref16,ref17,ref18}. These innovations highlight DL's ability to capture temporal dependencies and complex properties in sequential data.

In computer vision, architectures such as ResNet \cite{ref19} and Vision Transformers (ViT) \cite{ref20} have revolutionized image classification, object detection, and semantic segmentation, achieving state-of-the-art accuracy while enabling applications that were previously unattainable.

In NLP, DL models such as BERT \cite{ref21} and GPT-3 \cite{ref22} have set new standards in tasks such as language translation, text summarization, and question answering, surpassing traditional methods and enabling more nuanced understanding and generation of human language \cite{ref23,ref24,ref25,ref26,ref27}. These models excel in NLP tasks due to their ability to capture contextual relationships within text using self-attention mechanisms. They are particularly effective at modeling semantic intricacies in textual data, enabling strong performance in complex classification tasks. However, these methods require not only substantial computational resources for fine-tuning and inference but also access to massive amounts of high-quality data to achieve their full potential. This reliance on extensive datasets and powerful hardware often makes them less practical for small-scale or resource-constrained projects, where data availability and computational capacity may be limited.

To address these challenges, this study integrates the Synthetic Minority Oversampling Technique (SMOTE) \cite{ref28} with Tomek Links (SMOTE-Tomek) \cite{ref29}, providing a robust preprocessing solution for imbalanced datasets. SMOTE oversamples minority classes by generating synthetic samples through interpolation, effectively enhancing the diversity of underrepresented classes while mitigating overfitting, a common issue with random oversampling methods. Complementing this, Tomek Links identifies and removes noisy or borderline samples, improving class separability by reducing overlapping data points between classes. This dual-action preprocessing strategy results in cleaner and more balanced training datasets, enabling ML models to better learn decision boundaries for minority classes.

The proposed approach is particularly suited to text-based datasets such as the PROMISE dataset, where class imbalance and noise frequently undermine classification performance. By addressing these issues, SMOTE-Tomek creates a foundation for training classical ML models, which can achieve useful performance without the computational demands of deep learning approaches. This study builds on the strengths of SMOTE-Tomek by integrating it into a stratified K-fold cross-validation framework, preserving the integrity of validation folds and ensuring rigorous evaluation.

In RE classification, SMOTE-Tomek addresses both class imbalance and noise, making it particularly suitable for the PROMISE dataset, where minority classes are significantly underrepresented. While previous studies have demonstrated the efficacy of SMOTE-Tomek for addressing class imbalance in the PROMISE dataset \cite{ref30}, this research extends its application by integrating it into a stratified K-fold cross-validation framework. This ensures that the validation folds remain untouched by resampling, preserving the original class distribution and enabling evaluation of the model's generalization capabilities. Furthermore, this study explores several classical ML models, evaluates their performance on the PROMISE dataset, and highlights the potential of lightweight solutions for efficient requirement classification.

This study makes the following key contributions:
\begin{itemize}[leftmargin=1.5em]
\item A systematic application of SMOTE-Tomek to address class imbalance in the PROMISE dataset, within a 10-fold cross-validation framework that ensures the validation set remains unaffected by resampling.
\item A comparative evaluation of classical ML models, demonstrating their potential for scalable and efficient requirement classification.
\item An analysis of Logistic Regression (LR) coefficients to interpret the relationship between features and requirement types, offering insights into the most frequent and influential words for each type.
\end{itemize}

The experimental results demonstrate that integrating SMOTE-Tomek with classical ML frameworks improves classification performance. LR, for instance, achieves an accuracy improvement from \pmv{59.85\%}{2.52\%} to \pmv{76.16\%}{2.58\%} under cross-validation, an absolute gain of 16.31 percentage points.

The remainder of this paper is structured as follows: Section 2 outlines the methodology, including an overview of the dataset, preprocessing steps, the class imbalance challenge, the SMOTE-Tomek approach, and the classical ML models utilized. Section 3 presents the results and their implications, and Section 4 concludes with directions for future research.

\section{Learning Method}
This section provides an overview of the methodology employed in this study, detailing the dataset, class imbalance challenges, and the preprocessing techniques applied to address them. It outlines the evaluated ML models, the implementation of the SMOTE-Tomek approach for data balancing, and the use of K-fold cross-validation to support robust performance assessment.

\subsection{Dataset}
\begin{figure}[H]
\centering
\includegraphics[width=0.82\textwidth]{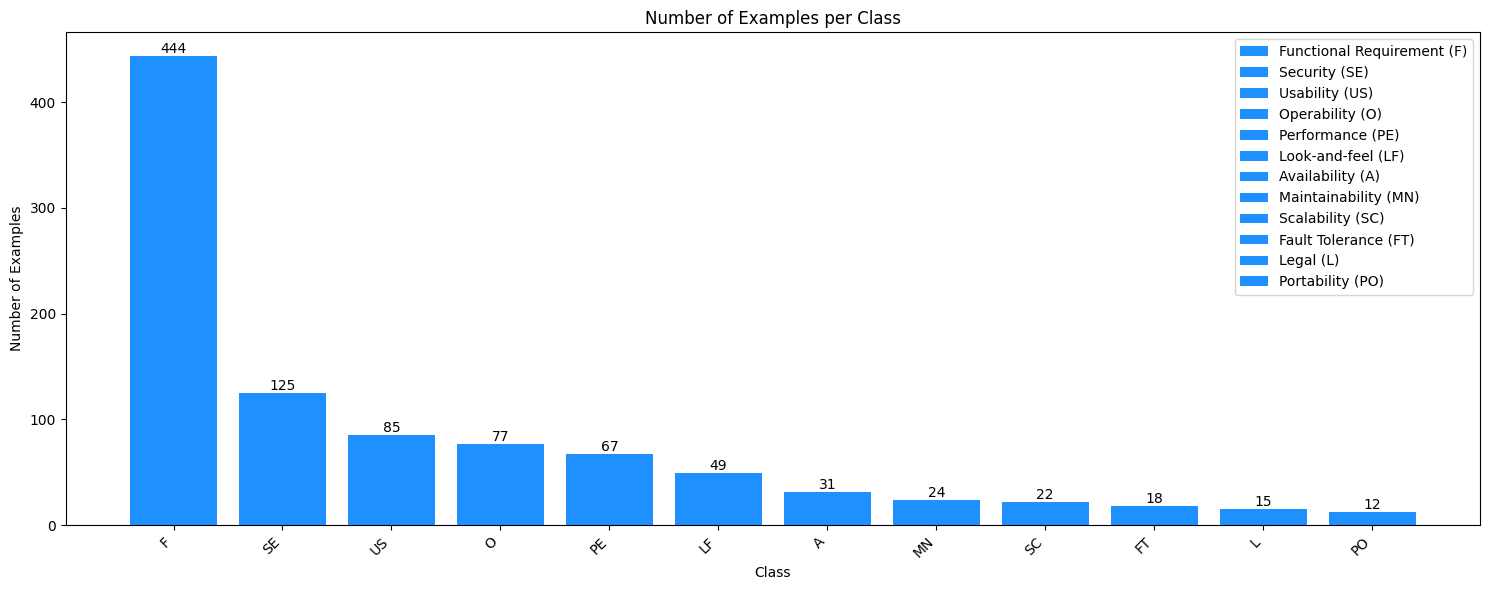}
\caption{Requirement type distribution in the PROMISE dataset.}
\label{fig:distribution}
\end{figure}

The study utilizes the expanded PROMISE dataset, a diverse collection of 969 software requirements extracted from Software Requirements Specification (SRS) documents using the Google search engine \cite{ref13}. The dataset contains 444 functional requirements (45.8\%) and 525 non-functional requirements (54.2\%), distributed across 12 distinct categories. These categories represent a spectrum of requirements in software engineering, such as Security (SE), Usability (US), Portability (PO), and Performance (PE). Despite its comprehensiveness, the dataset exhibits notable class imbalance. For example, the Portability category is severely underrepresented, comprising only 12 requirements (1.24\%), while the most prevalent non-functional class, Security, includes 125 requirements (12.9\%). This disparity affects the performance and generalizability of ML models. Figure~\ref{fig:distribution} illustrates the distribution of the dataset across all requirement types.

The following is a concise description of the requirement types used in this study:

\begin{description}[style=multiline,leftmargin=4.5cm,labelwidth=4.2cm,font=\normalfont\bfseries]
\item[Functional Requirement (F)] Defines the specific functionalities and behaviors the system must perform to meet stakeholder needs.
\item[Availability Requirement (A)] Ensures that the system remains operational and accessible under defined conditions and constraints.
\item[Legal Requirement (L)] Addresses compliance with legal, regulatory, and contractual obligations relevant to the system's deployment and use.
\item[Look-and-Feel Requirement (LF)] Specifies the aesthetic and user interface characteristics to align with user expectations and branding.
\item[Maintainability Requirement (MN)] Focuses on the system's ability to be easily modified, updated, or repaired during its lifecycle.
\item[Operability Requirement (O)] Ensures that the system can be operated efficiently and effectively within its intended environment.
\item[Performance Requirement (PE)] Defines the system's capability to deliver expected throughput, response time, and efficiency under specified conditions.
\item[Scalability Requirement (SC)] Addresses the system's ability to adapt to increased workloads or expanded functionalities without degradation.
\item[Security Requirement (SE)] Protects the system and its data against unauthorized access, breaches, and vulnerabilities.
\item[Usability Requirement (US)] Ensures the system is intuitive and user-friendly, enabling stakeholders to interact with it effectively.
\item[Fault Tolerance Requirement (FT)] Ensures the system can continue functioning correctly even in the event of hardware or software failures.
\item[Portability Requirement (PO)] Specifies the system's ability to be deployed or operated across various platforms, environments, or configurations.
\end{description}

\begin{table}[H]
\centering
\caption{Example requirements and their corresponding types from the PROMISE dataset}
\label{tab:examples}
\small
\begin{tabularx}{\textwidth}{Y L{0.27\textwidth}}
\toprule
\textbf{Requirement Example} & \textbf{Requirement Type} \\
\midrule
The product shall have audit capabilities. The product shall store messages for a minimum of one year for audit and transaction tracking purposes. & Security (SE) \\
The recycled parts search results provided to the estimator shall be retrieved by the system. & Functional (F) \\
Users should be able to access their streaming movies in under two clicks after logging into the website. & Usability (US) \\
The product shall be available 24 hours per day, seven days per week. & Availability (A) \\
The system shall help the user avoid making mistakes while scheduling classes and clinicals for the nursing students. & Usability (US) \\
The system shall display a detailed invoice for the current order once it is confirmed. & Functional (F) \\
The system will use a secured database. & Security (SE) \\
The leads washing functionality will store any potential lead duplicates returned by the enterprise system. & Functional (F) \\
The system shall interface with the faculty central server. & Operability (O) \\
The product shall be internet browser independent. The product must run using Internet Explorer and Netscape Navigator. & Maintainability (MN) \\
\bottomrule
\end{tabularx}
\end{table}

\subsection{Problem Definition: Multiclass Classification}
While some studies in requirements engineering focus on a binary classification task, distinguishing only between functional and non-functional requirements, the present study addresses a more complex multiclass classification problem.

The objective is to automatically categorize each requirement into one of the 12 distinct classes available in the PROMISE dataset, comprising Functional plus 11 non-functional subtypes. This formulation is more challenging due to the extreme class imbalance, where minority classes such as Portability or Fault Tolerance represent less than 2\% of the dataset. By defining the task as multiclass, we aim to demonstrate the robustness of the SMOTE-Tomek approach in learning fine-grained semantic boundaries across a diverse spectrum of requirement types, rather than identifying only broad categories.

\subsection{Pre-Processing}
Transforming unstructured text into a numerical representation suitable for ML algorithms is a fundamental preprocessing step in many NLP tasks. The PROMISE dataset comprises individual sentences rather than full documents. Term Frequency-Inverse Document Frequency (TF-IDF) remains a highly effective vectorization technique \cite{ref31,ref32}. TF-IDF quantifies the importance of a term within a sentence while considering its prevalence across the entire dataset of sentences. By emphasizing terms that are frequent within a specific sentence but rare across the dataset, TF-IDF captures features that are contextually meaningful and relevant to the classification task. Despite the brevity of the textual units, TF-IDF preserves the semantic significance of terms.

\subsection{Class Imbalance Challenge}
Class imbalance poses a critical challenge in training ML models, as they tend to prioritize majority classes, resulting in poor generalization and suboptimal performance on minority classes \cite{ref33}. Such biases can affect the usability and reliability of classification models in practical applications, where correctly identifying minority classes is often essential. Balancing the training set helps the model give greater attention to underrepresented classes and reduces bias toward majority classes.

\subsection{SMOTE-Tomek}
Among various methods to address data imbalance, SMOTE-Tomek is a hybrid technique that combines oversampling and data cleaning. SMOTE addresses the imbalance by generating synthetic samples for minority classes, producing a more equitable class distribution in the training data. However, SMOTE alone can introduce noise by oversampling borderline or overlapping samples. Tomek Links complement SMOTE by identifying and removing borderline or ambiguous samples that may hinder model training. Together, SMOTE-Tomek enhances data quality while balancing the class distribution during training. Unlike random oversampling, which duplicates existing minority class samples, SMOTE creates new synthetic samples based on linear interpolation between existing samples.

In NLP applications, where textual data can be complex and imbalanced, the synthetic examples generated by SMOTE are not direct textual entities. They are feature vectors representing characteristics of the minority class, as illustrated in Table~\ref{tab:resampled}. These vectors serve as abstract representations, enabling the model to learn structural properties of underrepresented classes without requiring additional annotated textual data.

The SMOTE-Tomek process is described in the following steps:
\begin{enumerate}[leftmargin=1.8em,itemsep=0.20em]
\item \textbf{Generate Synthetic Samples for Minority Classes:} For each sample $x_i$ belonging to the minority class, synthetic samples are generated using linear interpolation:
\begin{equation}
 x_{\mathrm{synthetic}} = x_i + \lambda(x_{\mathrm{nn}}-x_i),
\end{equation}
where $x_{\mathrm{nn}}$ is a randomly selected nearest neighbor of $x_i$ from the same minority class, and $\lambda \in [0,1]$ is a random scalar.

\item \textbf{Identify Tomek Links:} A Tomek Link is a pair of samples $x_i$ and $x_j$ that are mutual nearest neighbors, meaning that $x_i$ is the closest sample to $x_j$ and vice versa. The two samples belong to different classes, with one from the minority class and the other from the majority class. The nearest-neighbor condition can be written as
\begin{equation}
 d(x_i,x_j)=\min_{x_k}\{d(x_i,x_k)\},
\end{equation}
with the reciprocal condition defined analogously for $x_j$.

\item \textbf{Remove Majority Class Samples:} For each identified Tomek Link, remove the sample belonging to the majority class to enhance class separation and reduce noise.
\end{enumerate}

\begin{table}[H]
\centering
\caption{Examples of resampled requirement representations using SMOTE-Tomek}
\label{tab:resampled}
\scriptsize
\begin{tabularx}{\textwidth}{Y L{0.23\textwidth}}
\toprule
\textbf{Requirement Representation} & \textbf{Requirement Type} \\
\midrule
based, established, lead, operate, physical, process, product, run, server, service, shall, structure, washing, web & Operability (O) \\
changes, edit, information, just, modify, normal, personal, read, securely, server, transmission, transmitted, users & Security (SE) \\
entry, shown & Functional (F) \\
initiator, mediator, messages, request, response, shall, user & Functional (F) \\
accommodate, compensatory, data, failures, fault, items, large, product, recovery, robust, routing, shall, technique, tolerance, transaction, using & Fault Tolerance (FT) \\
explanatory, intuitive, self, shall & Usability (US) \\
control, details, employees, site, view & Functional (F) \\
browsers, ce, compatible, environments, expected, interface, major, operating, order, palm, product, run, standards, systems, usable, variety, web, wide, windows & Portability (PO) \\
800x600, 95, appropriately, correct, corrected, display, feel, higher, implementation, incorrect, intranet, look, notification, pages, prior, provide, remaining, resolutions, shall, uniform, web, week & Look-and-Feel (LF) \\
able, alongside, days, environment, function, installed, java, operating, product, runtime, server, shall, software & Operability (O) \\
\bottomrule
\end{tabularx}
\end{table}

\subsection{Stratified K-fold Cross-Validation Method}
Stratified K-fold cross-validation provides advantages over a single train-test split, particularly when classes are imbalanced. By dividing the dataset into $K$ folds and using each fold for validation exactly once, the method makes broad use of the available data. Stratified sampling maintains the proportional representation of classes in each fold, reducing the risk that minority classes are omitted from training or validation subsets.

\begin{figure}[H]
\centering
\includegraphics[width=0.78\textwidth]{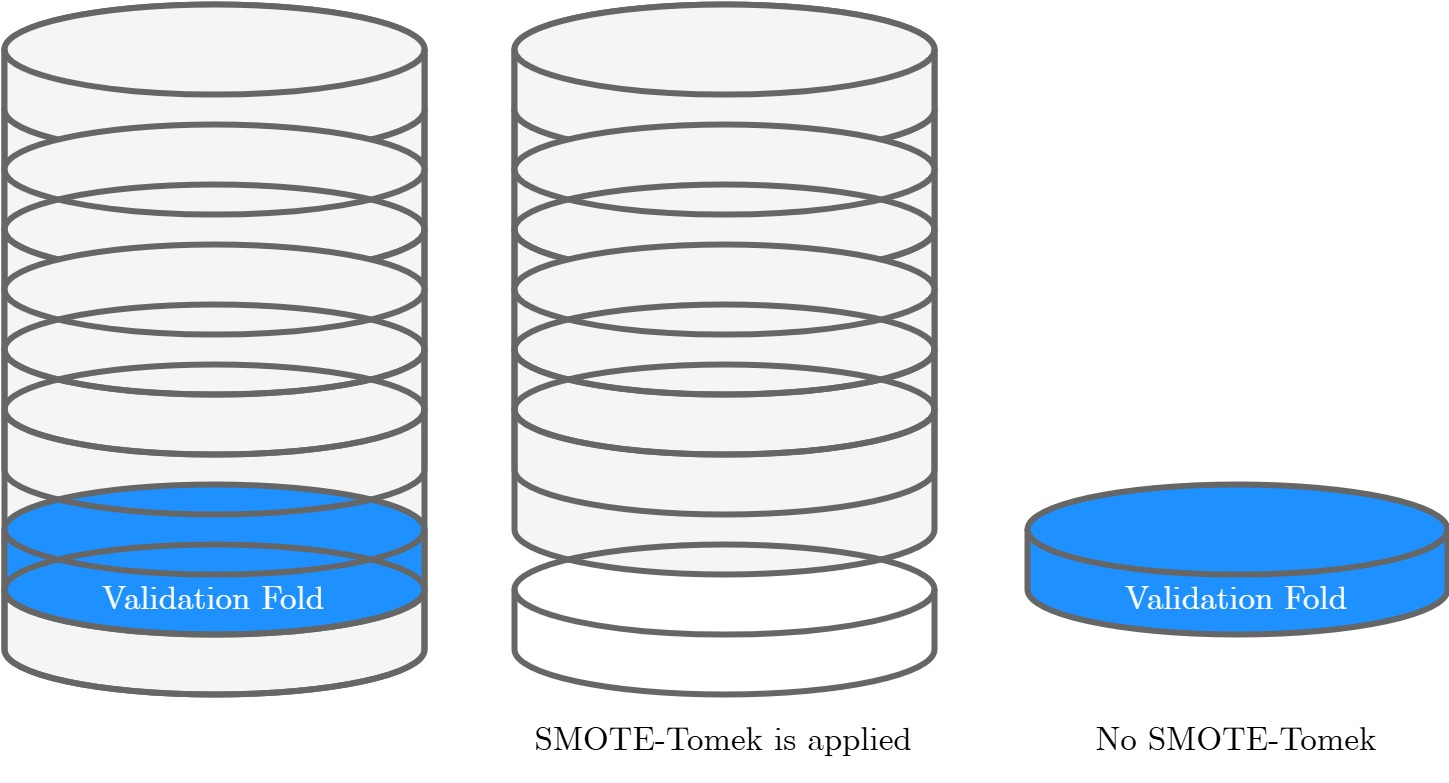}
\caption{Stratified cross-validation with SMOTE-Tomek applied only to the training folds; the validation fold is not resampled.}
\label{fig:cv}
\end{figure}

To maintain the integrity of the cross-validation process, SMOTE-Tomek resampling is applied solely to the training folds during each iteration of the 10-fold stratified cross-validation. Specifically, in each iteration, the resampling technique is applied to the nine training folds, while the validation fold remains untouched. This preserves the original distribution of the validation data, avoids data leakage from synthetic samples into validation, and enables a more realistic evaluation of model performance. As illustrated in Figure~\ref{fig:cv}, SMOTE-Tomek is applied only to the training folds.

\subsection{Training Algorithm}
The proposed method, outlined in Algorithm~\ref{alg:training}, integrates stratified K-fold cross-validation to preserve class proportions across folds. SMOTE-Tomek is applied solely to the training folds to prevent data leakage and preserve the integrity of validation metrics. The algorithm also includes final training on the complete balanced dataset after cross-validation.

\begin{table}[H]
\centering
\caption{Notations and definitions}
\label{tab:notation}
\small
\begin{tabular}{L{0.18\textwidth} L{0.68\textwidth}}
\toprule
\textbf{Notation} & \textbf{Definition} \\
\midrule
$\mathcal{D}$ & Complete dataset used for training and evaluation \\
$K$ & Number of folds in stratified K-fold cross-validation \\
$\mathcal{F}_i$ & Fold $i$ used as the validation set in the $i$-th iteration \\
$\mathcal{D}_{\mathrm{train}}$ & Training set formed by $K-1$ folds \\
$\mathcal{D}_{\mathrm{val}}$ & Validation set corresponding to fold $i$ \\
$\mathcal{M}_i$ & Model trained in the $i$-th cross-validation iteration \\
\bottomrule
\end{tabular}
\end{table}

\begin{algorithmbox}{Training with SMOTE-Tomek and Stratified K-Fold Cross-Validation}\label{alg:training}
\textbf{Require:} $\mathcal{D}$, $K$, $\mathcal{M}$
\begin{enumerate}[leftmargin=2em,label=\arabic*:,itemsep=0.10em,topsep=0.18em]
\item Split $\mathcal{D}$ into $K$ stratified folds $\{\mathcal{F}_1,\mathcal{F}_2,\ldots,\mathcal{F}_K\}$, preserving class proportions.
\item For fold $i=1$ to $K$:
  \begin{enumerate}[leftmargin=2.1em,label*=\arabic*.,itemsep=0.08em,topsep=0.10em]
  \item $\mathcal{D}_{\mathrm{train}} \leftarrow \bigcup_{j\neq i}\mathcal{F}_j$.
  \item $\mathcal{D}_{\mathrm{val}} \leftarrow \mathcal{F}_i$.
  \item $\mathcal{D}_{\mathrm{train,balanced}} \leftarrow$ apply SMOTE-Tomek only to $\mathcal{D}_{\mathrm{train}}$.
  \item Train $\mathcal{M}_i$ on $\mathcal{D}_{\mathrm{train,balanced}}$.
  \item Evaluate $\mathcal{M}_i$ on $\mathcal{D}_{\mathrm{val}}$ and compute the metrics.
  \end{enumerate}
\item Compute average metrics across all $K$ folds.
\item Train $\mathcal{M}_{\mathrm{final}}$ on $\mathcal{D}$ after applying SMOTE-Tomek to the complete training data.
\item Return $\mathcal{M}_{\mathrm{final}}$ and the average cross-validation metrics.
\end{enumerate}
\end{algorithmbox}

\subsection{Classical ML Models}
The following ML models were evaluated in this study to capture diverse classification perspectives:
\begin{itemize}[leftmargin=1.5em]
\item \textbf{Decision Tree (DT):} A tree-based algorithm that splits the data into subsets based on feature values, creating a hierarchy of decisions to classify data.
\item \textbf{Random Forest:} An ensemble method combining multiple decision trees, where each tree votes and the majority decision is taken.
\item \textbf{Support Vector Machine (SVM) with Linear Kernel:} A linear classifier that finds the hyperplane maximizing the margin between classes.
\item \textbf{SVM with RBF Kernel:} A nonlinear classifier that maps data into a higher-dimensional space using the radial basis function kernel.
\item \textbf{Naive Bayes:} A probabilistic model based on Bayes' theorem, assuming feature independence, and commonly used for text classification.
\item \textbf{Logistic Regression:} A statistical model that predicts class probabilities using the logistic function.
\item \textbf{K-Nearest Neighbors (KNN):} A non-parametric algorithm that classifies samples based on the majority class of their nearest neighbors in feature space.
\item \textbf{Gradient Boosting:} An iterative ensemble method that builds sequential trees, correcting errors of previous trees to minimize classification error.
\item \textbf{AdaBoost:} An adaptive boosting algorithm that combines weak classifiers and gives higher weights to misclassified samples.
\item \textbf{CatBoost:} A gradient boosting algorithm optimized for categorical features.
\end{itemize}

\section{Results and Discussion}
This section evaluates the performance of the ML models for requirement type classification, with a focus on the impact of SMOTE-Tomek preprocessing. It outlines the error metrics, compares baseline performance with the performance obtained after SMOTE-Tomek integration, highlights the interpretability of Logistic Regression, and discusses the observed improvements and limitations.

\subsection{Error Metrics}
Let TP, FP, TN, and FN denote true positives, false positives, true negatives, and false negatives, respectively. Precision quantifies the proportion of correctly predicted positive cases among all predicted positive cases:
\begin{equation}
\mathrm{Precision}=\frac{TP}{TP+FP}.
\end{equation}
Recall evaluates the proportion of actual positive cases correctly identified:
\begin{equation}
\mathrm{Recall}=\frac{TP}{TP+FN}.
\end{equation}
The F1-score is the harmonic mean of precision and recall:
\begin{equation}
\mathrm{F1}=\frac{2TP}{2TP+FP+FN}.
\end{equation}
Accuracy assesses overall correctness:
\begin{equation}
\mathrm{Accuracy}=\frac{TP+TN}{TP+TN+FP+FN}.
\end{equation}
Matthews Correlation Coefficient (MCC) provides a summary of classification quality under imbalance. For a binary confusion matrix, it is defined as
\begin{equation}
\mathrm{MCC}=\frac{TP\cdot TN-FP\cdot FN}{\sqrt{(TP+FP)(TP+FN)(TN+FP)(TN+FN)}}.
\end{equation}
The reported experiment uses the corresponding multiclass extension. MCC ranges from $-1$ to $+1$, where $+1$ represents perfect classification, $0$ indicates random or uninformed classification, and $-1$ represents complete disagreement.

\subsection{Baseline Performance Without SMOTE-Tomek}
\begin{figure}[H]
\centering
\includegraphics[width=0.89\textwidth]{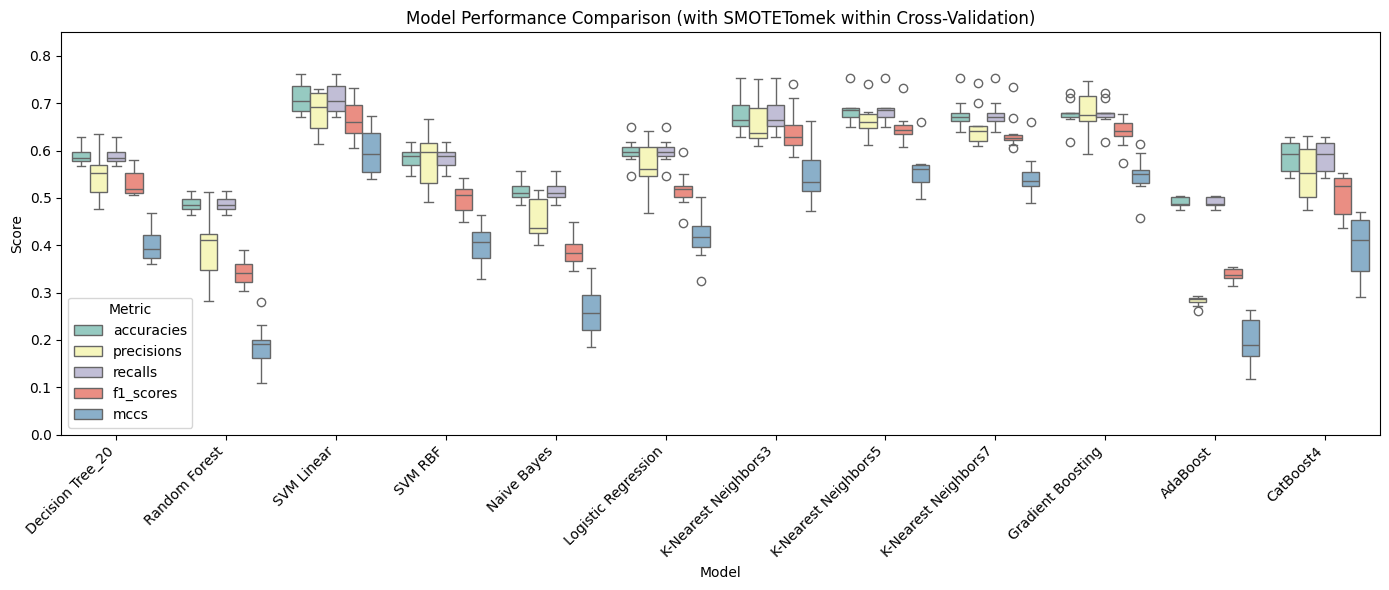}
\caption{Cross-validation performance of machine learning models without SMOTE-Tomek preprocessing.}
\label{fig:baseline}
\end{figure}

The baseline experimental results are summarized in Table~\ref{tab:baseline} and visualized in Figure~\ref{fig:baseline}. Among all evaluated classifiers, the linear SVM achieved the strongest overall performance, attaining a mean accuracy of 71.10\%, precision of 68.25\%, recall of 71.10\%, F1-score of 66.47\%, and an MCC of 0.5977. This indicates a favorable balance between predictive accuracy and robustness under class imbalance.

KNN classifiers with $k=3$ and $k=5$ also demonstrated competitive performance, with mean accuracies of 67.69\% and 68.42\%, respectively, and MCC values close to that of the linear SVM. These results suggest that local neighborhood-based methods can capture discriminative structure in the PROMISE dataset despite class imbalance. Gradient Boosting achieved a mean accuracy of 66.98\% and an MCC of 0.5344.

In contrast, Random Forest and AdaBoost showed weaker results. AdaBoost, in particular, produced a mean precision of 28.21\% and an MCC of 0.1963. Logistic Regression and Decision Trees achieved moderate performance but were outperformed by SVM and KNN variants without imbalance-aware preprocessing.

\begin{table}[H]
\centering
\caption{Performance for each model without SMOTE-Tomek}
\label{tab:baseline}
\scriptsize
\resizebox{\textwidth}{!}{%
\begin{tabular}{lccccc}
\toprule
\textbf{Model} & \textbf{Mean Accuracy (\%)} & \textbf{Mean Precision (\%)} & \textbf{Mean Recall (\%)} & \textbf{Mean F1-Score (\%)} & \textbf{Mean MCC} \\
\midrule
Decision Tree       & \pmv{59.55}{2.72} & \pmv{54.71}{4.08} & \pmv{59.55}{2.72} & \pmv{53.48}{3.15} & \pmv{0.4116}{0.0473} \\
Random Forest       & \pmv{48.71}{1.48} & \pmv{38.26}{8.61} & \pmv{48.71}{1.48} & \pmv{34.11}{2.53} & \pmv{0.1796}{0.0672} \\
SVM (Linear)        & \pmv{71.10}{3.09} & \pmv{68.25}{4.18} & \pmv{71.10}{3.09} & \pmv{66.47}{3.73} & \pmv{0.5977}{0.0456} \\
SVM (RBF)           & \pmv{58.31}{2.05} & \pmv{58.05}{5.33} & \pmv{58.31}{2.05} & \pmv{49.74}{2.88} & \pmv{0.4010}{0.0382} \\
Naive Bayes         & \pmv{51.49}{2.04} & \pmv{45.57}{3.96} & \pmv{51.49}{2.04} & \pmv{38.80}{2.91} & \pmv{0.2566}{0.0510} \\
Logistic Regression & \pmv{59.85}{2.52} & \pmv{56.85}{5.00} & \pmv{59.85}{2.52} & \pmv{51.85}{3.65} & \pmv{0.4181}{0.0470} \\
KNN ($k=3$)         & \pmv{67.69}{3.84} & \pmv{66.11}{4.98} & \pmv{67.69}{3.84} & \pmv{64.34}{4.55} & \pmv{0.5478}{0.0586} \\
KNN ($k=5$)         & \pmv{68.42}{2.69} & \pmv{66.37}{3.29} & \pmv{68.42}{2.69} & \pmv{64.95}{3.12} & \pmv{0.5564}{0.0427} \\
KNN ($k=7$)         & \pmv{67.80}{2.93} & \pmv{65.07}{3.98} & \pmv{67.80}{2.93} & \pmv{63.84}{3.57} & \pmv{0.5468}{0.0452} \\
Gradient Boosting   & \pmv{66.98}{2.92} & \pmv{66.37}{5.05} & \pmv{66.98}{2.92} & \pmv{63.13}{3.05} & \pmv{0.5344}{0.0446} \\
AdaBoost            & \pmv{49.02}{1.14} & \pmv{28.21}{0.94} & \pmv{49.02}{1.14} & \pmv{33.71}{1.33} & \pmv{0.1963}{0.0472} \\
CatBoost            & \pmv{58.82}{3.11} & \pmv{55.21}{5.54} & \pmv{58.82}{3.11} & \pmv{50.73}{4.32} & \pmv{0.3978}{0.0636} \\
\bottomrule
\end{tabular}}
\end{table}

\subsection{Enhanced Performance Using SMOTE-Tomek}
\begin{figure}[H]
\centering
\includegraphics[width=0.89\textwidth]{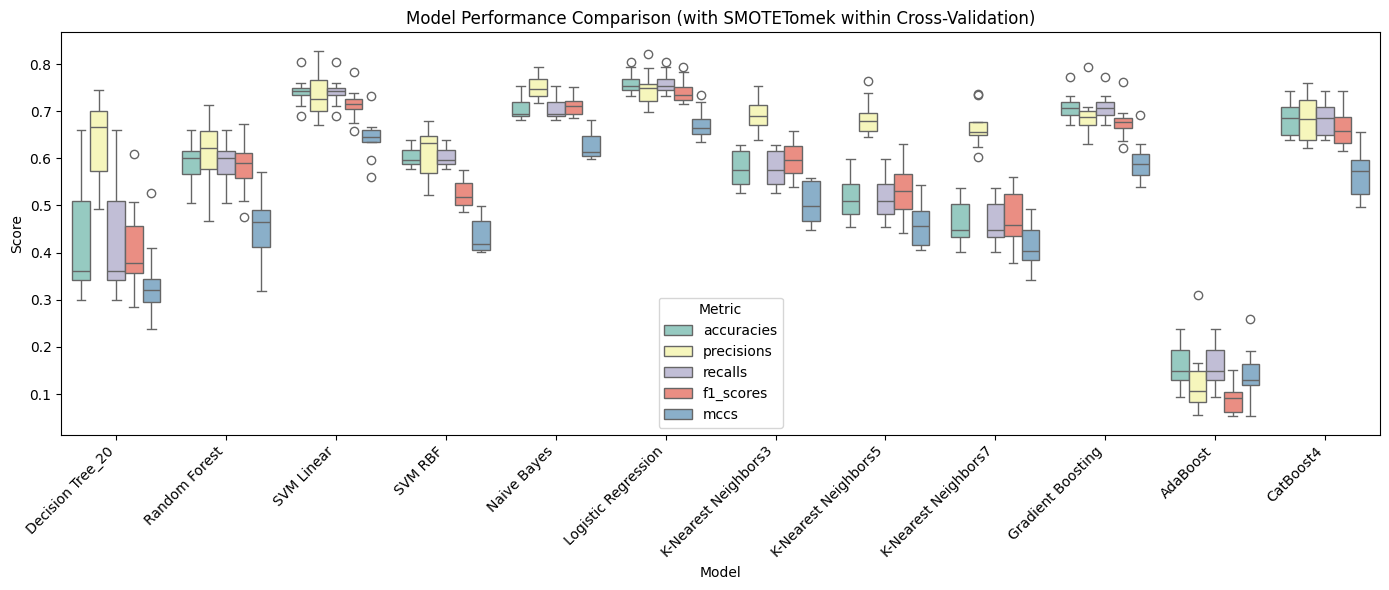}
\caption{Cross-validation performance of machine learning models with SMOTE-Tomek preprocessing.}
\label{fig:smote}
\end{figure}

The impact of SMOTE-Tomek preprocessing on classification performance is presented in Table~\ref{tab:smote} and Figure~\ref{fig:smote}. Following the application of SMOTE-Tomek, Logistic Regression exhibited the most pronounced improvement across the reported evaluation metrics, achieving a mean accuracy of \pmv{76.16\%}{2.58\%}, precision of \pmv{75.31\%}{4.30\%}, recall of \pmv{76.16\%}{2.58\%}, F1-score of \pmv{74.34\%}{2.93\%}, and an MCC of \pmv{0.6736}{0.0370}.

Relative to the baseline Logistic Regression results in Table~\ref{tab:baseline}, this corresponds to an absolute accuracy gain of 16.31 percentage points. MCC increased from \pmv{0.4181}{0.0470} to \pmv{0.6736}{0.0370}, an absolute gain of 0.2555. Performance was further improved by approximately one percentage point through hyperparameter optimization using $C=10$, an L2 regularization penalty, and the SAGA solver \cite{ref34}.

Several additional classifiers benefited from SMOTE-Tomek preprocessing. Naive Bayes and Gradient Boosting achieved mean accuracies exceeding 70\%, with MCC values of 0.6286 and 0.5938, respectively. In contrast, AdaBoost remained highly sensitive to the resampled data distribution, achieving a mean accuracy of 16.10\% and an MCC of 0.1422.

\begin{table}[H]
\centering
\caption{Performance for each model with SMOTE-Tomek}
\label{tab:smote}
\scriptsize
\resizebox{\textwidth}{!}{%
\begin{tabular}{lccccc}
\toprule
\textbf{Model} & \textbf{Mean Accuracy (\%)} & \textbf{Mean Precision (\%)} & \textbf{Mean Recall (\%)} & \textbf{Mean F1-Score (\%)} & \textbf{Mean MCC} \\
\midrule
Decision Tree       & \pmv{41.79}{12.44} & \pmv{63.89}{7.90} & \pmv{41.79}{12.44} & \pmv{40.43}{9.53} & \pmv{0.3363}{0.0780} \\
Random Forest       & \pmv{59.03}{4.50}  & \pmv{60.82}{6.96} & \pmv{59.03}{4.50}  & \pmv{58.05}{5.39} & \pmv{0.4562}{0.0686} \\
SVM (Linear)        & \pmv{74.20}{2.83}  & \pmv{73.46}{4.58} & \pmv{74.20}{2.83}  & \pmv{71.44}{3.26} & \pmv{0.6428}{0.0424} \\
SVM (RBF)           & \pmv{60.27}{2.12}  & \pmv{61.34}{5.40} & \pmv{60.27}{2.12}  & \pmv{52.68}{3.09} & \pmv{0.4358}{0.0346} \\
Naive Bayes         & \pmv{70.48}{2.28}  & \pmv{75.04}{2.42} & \pmv{70.48}{2.28}  & \pmv{71.17}{2.06} & \pmv{0.6286}{0.0280} \\
Logistic Regression & \pmv{76.16}{2.58}  & \pmv{75.31}{4.30} & \pmv{76.16}{2.58}  & \pmv{74.34}{2.93} & \pmv{0.6736}{0.0370} \\
KNN ($k=3$)         & \pmv{57.79}{3.65}  & \pmv{69.21}{3.19} & \pmv{57.79}{3.65}  & \pmv{59.59}{3.78} & \pmv{0.5063}{0.0413} \\
KNN ($k=5$)         & \pmv{51.60}{4.31}  & \pmv{68.56}{3.76} & \pmv{51.60}{4.31}  & \pmv{53.00}{5.20} & \pmv{0.4596}{0.0451} \\
KNN ($k=7$)         & \pmv{46.34}{4.45}  & \pmv{66.64}{4.05} & \pmv{46.34}{4.45}  & \pmv{47.11}{5.69} & \pmv{0.4148}{0.0463} \\
Gradient Boosting   & \pmv{70.90}{2.77}  & \pmv{69.08}{4.15} & \pmv{70.90}{2.77}  & \pmv{67.76}{3.56} & \pmv{0.5938}{0.0419} \\
AdaBoost            & \pmv{16.10}{4.42}  & \pmv{12.67}{6.97} & \pmv{16.10}{4.42}  & \pmv{9.06}{3.07}  & \pmv{0.1422}{0.0526} \\
CatBoost            & \pmv{68.21}{3.32}  & \pmv{68.34}{4.55} & \pmv{68.21}{3.32}  & \pmv{66.37}{3.83} & \pmv{0.5633}{0.0492} \\
\bottomrule
\end{tabular}}
\end{table}

\subsection{Feature Analysis and Model Interpretability}
One of the primary advantages of linear models such as Logistic Regression is their interpretability. LR allows direct examination of feature importance through learned coefficients. To investigate the effect of SMOTE-Tomek preprocessing on the model's learned representation, we analyzed the top word coefficients for each requirement type in the multiclass classification setting.

\begin{figure}[H]
\centering
\includegraphics[width=0.91\textwidth]{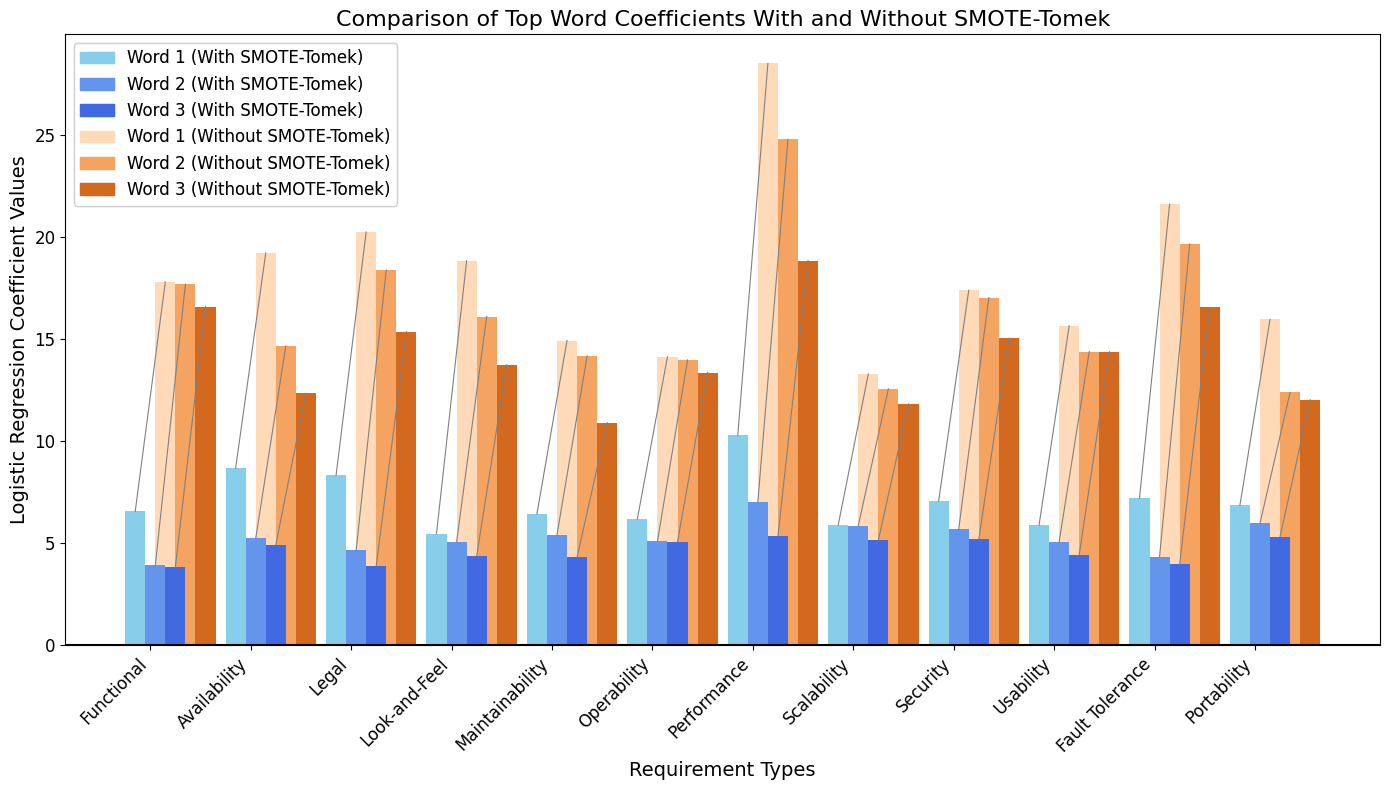}
\caption{Comparison of top word coefficients with and without SMOTE-Tomek preprocessing across requirement types.}
\label{fig:coefficients}
\end{figure}

Figure~\ref{fig:coefficients} visualizes the absolute values of the top three word coefficients for each requirement category, comparing the LR model trained on the imbalanced baseline dataset with the model trained using SMOTE-Tomek.

The analysis reveals a clear difference in coefficient magnitudes. Without SMOTE-Tomek, represented by the orange shades, the model exhibits large and variable coefficient values, particularly for minority classes such as Performance, Legal, and Fault Tolerance. For example, the top word for the Performance class reaches a coefficient value near 30. This pattern suggests that the baseline model may rely heavily on a small number of terms in classes with limited training examples.

With SMOTE-Tomek, represented by the blue shades, the top coefficient magnitudes are lower, mostly between 4 and 10, and are more similar across the three displayed features within each class. The result is consistent with reduced reliance on isolated terms and a more distributed lexical representation. Because the figure compares coefficient magnitudes rather than formal stability estimates across repeated runs, these observations should be interpreted descriptively.

\subsection{Discussion}
The highest observed accuracy was 76.16\% for Logistic Regression after applying SMOTE-Tomek. This result highlights both the potential and the limitations of classical ML models for requirements classification. Although performance improved relative to the Logistic Regression baseline, data scarcity and class imbalance remain important constraints in the PROMISE dataset. These challenges motivate the use of larger and more representative datasets to capture the diversity and nuances of requirement types.

Stratified K-fold cross-validation supported an evaluation framework that preserved the original class distribution in every validation fold. Crucially, the validation folds remained untouched by resampling. This protocol reduces the risk of data leakage from synthetic samples and highlights the importance of rigorous evaluation when testing class-imbalance mitigation strategies.

SMOTE-Tomek was associated with a substantial improvement in Logistic Regression metrics. Accuracy increased from \pmv{59.85\%}{2.52\%} to \pmv{76.16\%}{2.58\%}. MCC increased from \pmv{0.4181}{0.0470} to \pmv{0.6736}{0.0370}, an absolute gain of 0.2555. LR benefited from the changed training distribution and remained interpretable and scalable for the classification task.

Despite the overall improvement, several challenges emerged:
\begin{itemize}[leftmargin=1.5em]
\item Ensemble methods such as AdaBoost and Random Forest underperformed even with SMOTE-Tomek preprocessing. AdaBoost achieved a mean accuracy of 16.10\% and an MCC of 0.1422.
\item Gradient Boosting and Linear SVM demonstrated competitive results, although their performance gains were more modest than that of LR.
\item The PROMISE dataset remains small and constrained in its coverage of requirement types. The restricted variety of classes and reliance on textual features alone limit generalization.
\end{itemize}

\section{Conclusions}
We demonstrated the potential of integrating SMOTE-Tomek preprocessing with classical ML models to address class imbalance in requirements classification. Logistic Regression exhibited the most substantial improvement, achieving \pmv{76.16\%}{2.58\%} accuracy compared with its baseline of \pmv{59.85\%}{2.52\%}. Its MCC increased from \pmv{0.4181}{0.0470} to \pmv{0.6736}{0.0370}, an absolute gain of 0.2555. The interpretability of LR enabled an analysis of feature importance, while the coefficient comparison indicated lower and more evenly distributed top coefficient magnitudes after SMOTE-Tomek preprocessing.

These findings emphasize the practicality and scalability of classical ML models for imbalanced text datasets, particularly in resource-constrained environments. The methodology may also be applicable to domains such as legal or healthcare document analysis, where similar imbalance challenges exist. Future work will explore larger datasets, advanced feature representations, and hybrid approaches combining lightweight models with deep learning techniques to improve performance and generalizability.

\section{Conflict of Interest Statement}
The author declares no conflict of interest related to this work.

\bibliographystyle{unsrt}
\bibliography{references}

\end{document}